\documentclass[]{aa}  

\usepackage{graphicx}
\usepackage{txfonts}
\usepackage{longtable}
\usepackage{lscape} % Tables in landscape orientation
\usepackage{multirow, array}
\usepackage{threeparttable} % - add notes under the Table
\usepackage{supertabular,booktabs}
\usepackage{float}
\usepackage{soul}
\usepackage{ulem}

\usepackage[urlcolor=cyan, colorlinks=true, citecolor=blue, linkcolor=blue]{hyperref}

\newcommand{\mysp}{}\def\mysp/{}

\begin{document}

   \title{Binary central stars of planetary nebulae in the Large Magellanic Cloud}
%   \titlerunning{Binary central stars}

   \author{M. G{\l}adkowski\inst{1}
          \and
          M. Hajduk\inst{2}
          \and
          R. Smolec\inst{3}
          \and
          R. Szczerba\inst{1}
          \and
          I. Soszy{\'n}ski\inst{4}
          }

   \institute{Nicolaus Copernicus Astronomical Center, Polish Academy of Sciences, ul. Rabia{\'n}ska 8, 87-100 Toru{\'n}, Poland
 %             \email{}
         \and
         Department of Geodesy,
         Institute of Geodesy and Civil Engineering,
         Faculty of Geoengineering,
         University of Warmia and Mazury, ul. Oczapowskiego 2, 10-719 Olsztyn, Poland
         \and
             Nicolaus Copernicus Astronomical Center, Polish Academy of Sciences, ul. Bartycka 18, 00-716 Warszawa, Poland        
         \and
             Astronomical Observatory, University of Warsaw, Al. Ujazdowskie 4, 00-478, Warszawa, Poland\\
         \email{seyfert@ncac.torun.pl}
             %\thanks{The university of heaven temporarily does not
              %       accept e-mails}
             }

   \date{Received --; accepted --}

% \abstract{}{}{}{}{}
% 5 {} token are mandatory
 
  \abstract
   {Close binary central stars of planetary nebulae (PNe) must have formed through a common envelope evolution during the giant phase experienced by one of the stars. Transfer of the angular momentum from the binary system to the envelope leads to the shortening of the binary separations from the radius of red giant to the radius of the order of few tenths of AU. Thus, close binary central stars of planetary nebulae are laboratories to study the common envelope phase of evolution. The close binary fraction in the Galaxy has been measured in various sky surveys, but the close binary fraction is not yet well constrained for the Magellanic Clouds, and our results may help the study of common envelope evolution in low-metallicity environments.}
   {This paper presents a continuation of our study of variability in the Magellanic Cloud planetary nebulae on the basis of data from the OGLE survey. Previously, we had analysed the OGLE data in the Small Magellanic Cloud. Here, the study is extended to the Large Magellanic Cloud (LMC). In this paper we search for close binary central stars with the aim to constrain the binary fraction and period distribution in the LMC. 
   }
   {We identified 290 counterparts of PNe in the LMC in the $I$-band images from the OGLE-III and OGLE-IV surveys. However, the light curves of ten objects were not accessible in the OGLE database, and thus we analysed the time series photometry of 280 PNe.
   }
   {In total, 32 variables were found, but 5 of them turned out to be foreground objects. Another 18 objects show irregular or regular variability that is not attributable to the binarity of their central stars. Their status and the nature of their variability will be verified in the follow-up paper. Nine binary central stars of PNe with periods between 0.24 and 23.6\,days were discovered. The obtained fraction for the LMC PNe is $3.3^{+2.6}_{-1.6}\%$ without correcting for incompleteness. This number is significantly lower than the 12--21\% derived in the analogous search in the Galactic bulge. We discuss this difference, taking into account observational biases. The lower binary fraction suggests a lower efficiency of the common envelope phase in producing close binaries in the LMC compared to the Galaxy.
   }
   {}


   \maketitle
%
%________________________________________________________________

%-------------------------------------------------------------
% ---------------------- INTRODUCTION ------------------------
%-------------------------------------------------------------
\section{Introduction}
\label{sec:introduction}

   The common envelope phase affects the evolution of many kinds of stars \citep{2013A&ARv..21...59I}, including SN\,Ia or binary neutron stars. Binarity also plays a vital role in the evolution of low- and intermediate-mass stars, which form the central stars of planetary nebulae (CSPNe) \citep{2020Sci...369.1497D}. During the common envelope episode the companion interacts with the envelope of the asymptotic giant branch (AGB) star. The binary pair loses orbital energy, which is transferred to the envelope. A close binary with a separation of $\leq 0.2 \, \rm{AU}$ forms from the initial binary, which originally had a separation up to the radius of the AGB star. The common envelope evolution leaves fingerprints in the form of morphological features (e.g. jets or fast outflows), which cannot be reconciled with the evolution of a single star \citep{2017NatAs...1E.117J}.

   Close binary nuclei are most frequently discovered through photometric variability. This variability might be caused by a few effects such as irradiation (i.e. the heating of a cold main-sequence component by the hot nucleus of a planetary nebula, PN), ellipsoidal variability, and eclipses. 

   The fraction of binary CSPNe in the Milky Way was established as 12--21\% on the basis of the third phase of the Optical Gravitational Lensing Experiment (OGLE-III) $I$-band photometry of 288 objects towards the Galactic bulge \citep{2009A&A...496..813M}. This is consistent with the 10--15\% fraction obtained by \cite{2000ASPC..199..115B}. \citet{2021MNRAS.506.5223J} achieved the somewhat higher binary fraction of 20.7\% using data from Kepler.
   
   Few attempts to search for variability in the PNe beyond the Milky Way have been made to date. \cite{2007apn4.confE..69S} presented their preliminary results of the analysis of $\sim$450 PNe in the Large Magellanic Cloud (LMC) using the data from the MACHO and SuperMACHO surveys. The authors tentatively detected three eclipsing variables, one with unusual variability in bipolar PN [RP2006]\,916 that was separately analysed in detail by \cite{2007ApJ...669L..25S}, and several dozen irregular variables. The brightness of each lobe of [RP2006]\,916 varies in different seasons of SuperMACHO observations, which may indicate the presence of binary CSPN. However, none of the eclipsing binaries have been confirmed so far. \cite{2007apn4.confE..69S} revealed only one eclipsing binary, [RP2006]\,1304, which has not been confirmed by us with a more sensitive and complete data set. \cite{2011A&A...531A.157M} analysed a sample of $\sim$100 PNe in the LMC on the basis of accessible near-infrared photometry from the VISTA Magellanic Cloud, and OGLE-III surveys. They found six periodic variables and six irregular ones; however, none of the periodic variables turned out to be a genuine PN. They have been classified as long-period variables, H{\sc ii} regions, and one symbiotic star candidate. \cite{2014A&A...561A...8H} used the OGLE-II and OGLE-III $I$-band data to study a sample of 52 CSPNe in the Small Magellanic Cloud (SMC), and they discovered the first binary CSPNe beyond the Milky Way with a period of 1.15 or 2.31\,days. This is the only binary CSPNe discovered beyond our Galaxy to date. 
   
   This paper is a continuation of our study of variability in the SMC PNe \citep{2014A&A...561A...8H}. The study is extended to the LMC. Through this study we expect to better understand the role of binarity in the formation and shaping of the PNe in the environments, which are characterised by different metallicities. We present OGLE-III and OGLE-IV $I$-band observations of the LMC, and report the discovery of nine binary nuclei. Another 18 variable LMC objects will be analysed in detail in a separate paper (G{\l}adkowski et al. in prep., hereafter Paper\,II).
   
   This paper is organised as follows. In Sect.~\ref{sec:pne_sample} we describe the sample and process of identifications of objects on the OGLE reference images. In Sect.~\ref{sec:observations} and \ref{sec:data_analysis} we present the observational details and the data analysis performed, while in Sect.~\ref{sec:notes} we present the light curve analysis details for each binary PN. Finally, Sect.~\ref{sec:discussion} discusses the results, and Sect.~\ref{sec:summary} presents a summary of our work.

%-------------------------------------------------------------
% ----------------------- PNE SAMPLE -------------------------
%-------------------------------------------------------------
\section{PNe sample}
\label{sec:pne_sample}

   Our sample of the LMC PNe is based on the list of 
   605 PNe from \citet{2014MNRAS.438.2642R}, which should 
   be relatively free of PNe mimics. However, variations in the background may lead to residual emission lines being mistaken for genuine emission, as discussed by \citep{2011A&A...531A.157M}. The objects were identified by visual comparison of finding charts available in the literature with the OGLE-III $I$-band reference images. In the case when the finding chart was not available in the literature, the identification was supported by the images ($B_{\rm j}$ and $R$ filters) from the VO-Paris MAMA (Machine Automatique \`a M\'esurer pour l’Astronomie) Atlas generated in the Aladin sky atlas (e.g. \citealt{1992ASSL..174..103G} at DOI: \href{https://doi.org/10.1007/978-94-011-2472-0_14}{10.1007/978-94-011-2472-0\_14}), and the coordinates of the object. When the identification was positive, and the OGLE-III coordinates of the object were determined, the identification on the OGLE-IV fields (if available) was made automatically. Using this approach we positively identified a total of 290 objects in the OGLE-III reference images out of 605 PNe. In spite of positive identifications, the light curves of ten objects were not accessible in the OGLE database because the photometric pipeline did not treat them as stars. Therefore, our full sample for the further light curve analysis consists of 280 objects. The full list of these objects is shown in Table~A.1, whereas Table~\ref{tab:pne_id_short} shows the information only for the first five objects from the entire list of PNe.
   
   Among the remaining 315 PNe the identification of 125 objects was ambiguous (usually two close stars), and thus the observational verification (e.g. with the $R$ or $I$, and  H$\alpha$ filters) is needed. Another 127 objects were located in the regions not covered by OGLE-III, but covered by the OGLE-IV fields, which are not publicly available yet. For 56 PNe no optical counterpart in the $I$-band was found. Finally, in the case of seven objects the names given in the online Supporting Information by \citet{2014MNRAS.438.2642R}\footnote{This online table is available at: \href{https://simbad.u-strasbg.fr/simbad/sim-ref?querymethod=bib\&simbo=on\&submit=submit\%20bibcode\&bibcode=2014MNRAS.438.2642R}{LINK}} were not recognisable by the Simbad database, and thus we could not find their coordinates and make identifications for them: [RP2006]\,469, [RP2006]\,3366, [RP2006]\,3876, [RP2006]\,3886, [RP2006]\,3923, [RP2006]\,4388, and [RP2006]\,4398.
   
   We cleaned the LMC PN sample of foreground Galactic objects that are H$\alpha$ emitters. We used Gaia DR3 parallaxes for this purpose \citep{2016A&A...595A...1G,2022arXiv220800211G}. The measured parallaxes should be 0.020\,mas for the LMC objects, assuming the distance given by \citet{2019Natur.567..200P}. It would be impossible to verify whether an individual object belongs to the LMC. The median parallax uncertainties are $0.02-0.03\,\mathrm{mas}$ for $\mathrm{G} < 15$, $0.07\,\mathrm{mas}$ at $\mathrm{G} = 17$, $0.5\,\mathrm{mas}$ at $\mathrm{G}=20$, and $1.3\,\mathrm{mas}$ at $\mathrm{G}=21$\,mag \citep{2022arXiv220800211G}. However, foreground mimics can have a measurable parallax if they are close enough.
   
   We cross-matched our PN candidates with the Gaia DR3 catalogue with the separations between Gaia and OGLE coordinates $\leq 5 \, \mathrm{arc sec}$. Positional accuracy of the two surveys is much better, but we aimed to include Galactic objects whose positions changed between the Gaia and OGLE observations. Then we corrected the Gaia coordinates for the proper motions between 2016.0 (Gaia DR3 epoch) and 2005 (approximate epoch of the OGLE-III photometric maps). We selected the objects with a separation (corrected for the epoch difference) of less than 0.5\,arcsec and parallax divided by its standard error $R_{\rm{Plx}} > 6$.
   
   In total, seven objects have measurable parallax (Table~\ref{tab:foreground}). Five of the foreground objects are variables. Their light curves show an artificial decrease in brightness because the photometry runs in a `fixed position mode' and is affected by high proper motion of the stars \citep{2002AcA....52..217U}. Six of these objects have well-determined astrometric solutions with the Renormalised Unit Weight Error (RUWE) indicator close to 1. The RUWE parameter $> 1.4$ for [RP2006]\,1649 (Table\,\ref{tab:foreground}) may indicate a binary system \citep{2022arXiv220605595G}.

%***************************************************************************************
%***************************************************************************************
%
%_____________________________________________________________
%                                             Two column Table 
%_____________________________________________________________
%
\begin{table*}
\footnotesize
\caption{List of the LMC PNe listed in \citet{2014MNRAS.438.2642R} identified in the OGLE-III and OGLE-IV fields (short version).}
          
\label{tab:pne_id_short}      
\centering          
\begin{tabular}{llccccccl} % 9 columns
\hline\hline                       
        &   \multicolumn{2}{c}{(J2000.0)}	\\\cmidrule{2-3}
Name    &   RA		&	Dec		&	$V$		&	$I$		&	$N_{\rm I}$	&    Ref &  Notes& Other names\\
        &   (h,m,s)	&	(d,m,s)	&	(mag)	&	(mag)	&				&        &       &			 \\
\hline
SMP\,LMC\,1      &	4 38 34.79	&	$-$70 36 43.4	&	15.131	&	16.286	&	373+587	&    [4] &  \ldots  &   LHA\,120-N\,182, LM\,1-2, WSPN\,1\\
{[RP2006]\,2419} &	4 40 33.45	&	$-$69 05 34.1	&	20.775	&	20.381	&	255+591	&    [7] &  \ldots  &   \ldots\\
{[M94b]\,6}      &	4 46 33.05	&	$-$70 12 47.0	&	19.567	&	20.453	&	424     &    [3] &  \ldots  &   LMC-MA\,3\\
SMP\,LMC\,5      &	4 48 08.56	&	$-$67 26 06.9	&	16.598	&	16.967	&	451+96	&    [4] &  \ldots  &   LHA\,20-S\,1\\
MGPN\,LMC\,2     &	4 48 32.50	&	$-$68 35 44.5	&	20.517	&	20.097	&	450+580	&    [2] &  \ldots  &   \ldots\\
\hline                  
\end{tabular}
\tablefoot{The table lists the names, coordinates (RA and Dec) for J2000, OGLE-III mean $V$ and $I$ magnitudes, number of measurements in the $I$-band in the OGLE-III $+$ OGLE-IV surveys (when available), reference for identification, notes, and other names. The full table is available in Appendix\,\ref{sec:app_pne_id}.}
\tablebib{[2]\,\citet{1992A&AS...92..571M}, [3]\,\citet{1994A&AS..103..235M}, [4]\,\citet{1997A&AS..121..407L}, [7]\,\citet{2013MNRAS.436..604R}.
}
\end{table*}
%***************************************************************************************
%***************************************************************************************

%***************************************************************************************
%***************************************************************************************
\begin{table*}
\footnotesize
\caption{List of foreground objects found in the sample.}             
\label{tab:foreground}      
\centering          
\begin{tabular}{lccccccccc} % 8 columns
\hline\hline                          
   Name		 &	    $V$  &	  $I$  	 &	Gaia DR3 number		 &	 Plx      &	ePlx     &	RPlx   & RUWE	 &	 $G$	 &	sep.  \\
                &   (mag)   &   (mag)   &                       &   (mas)    & (mas)    &         &         &   (mag)   &   (sec)\\
\hline
{[RP2006]\,958}	&	20.081	&	17.337	&  4658530024142565120	&	2.178	&	0.167	&	13.01     &	1.047	&	18.753	&	0.4441 \\
{[RP2006]\,445}	&	19.863	&	17.184	&  4651843657712891776	&	2.524	&	0.147	&	17.07     &	0.958	&	18.610	&	0.3021 \\
{[RP2006]\,492}	&	15.377	&	14.640	&  4651071040381784960	&	0.513	&	0.022	&	22.73     &	1.0     &	15.215	&	0.1038 \\
{[RP2006]\,524}	&	18.539	&	16.125	&  4651831563135888512	&	3.751	&	0.080	&	46.44     &	1.07	&	17.462	&	0.0513 \\
{[RP2006]\,857}	&	19.266	&	16.494	&  4658121693014774400	&	5.231	&	0.095	&	54.77     &	1.001	&	17.972	&	0.3312 \\
{[RP2006]\,4305}&	17.369	&	14.705	&  5278572432368397568	&	8.944	&	0.037	&	236.03    &	1.1     &	16.131	&	0.1243 \\
{[RP2006]\,1649}&   19.759  &   16.613  &  4652042252698588544  &   5.610   &   0.228   &   24.52     & 1.856   &  18.218   &   0.4959 \\
\hline                  
\end{tabular}
\tablefoot{The table lists the names, OGLE-III mean $V$ and $I$ magnitudes, Gaia DR3 numbers, parallaxes and the relevant error, RUWE parameter, Gaia $G$ magnitude, and separation between OGLE and Gaia coordinates.}

\end{table*}
%***************************************************************************************
%***************************************************************************************

%***************************************************************************************
%***************************************************************************************
\begin{table*}
\footnotesize
\caption{List and properties of binary central stars of PNe.}             
\label{tab:list_binary_pne}      
\centering          
\begin{tabular}{llcccccc} % 8 columns
\hline\hline                       
		&	\multicolumn{2}{c}{(J2000.0)}	\\\cmidrule{2-3}
Name	&	RA		&	Dec		&	$V$	    &	$I$		&	$P_{\rm 1}$	&	$P_{\rm 2}$ & Ref \\
		&	(h,m,s)	&	(d,m,s)	&	(mag)	&	(mag)	&	(days)		&	(days)      & \\
\hline
{[RP2006]\,1773} &	5 04 35.10	&	$-$70 06 19.1	&	18.876	&   18.263  &	6.0615  &	\ldots   & [1] \\
{[RP2006]\,1488} &	5 13 57.12	&	$-$68 25 09.0	&	20.439	&   19.706  &	0.5301	&	8.6002   & [1] \\
{[RP2006]\,1429} &	5 19 18.62	&	$-$68 49 12.1	&	19.975	&	19.078	&	6.0130	&   \ldots   & [1] \\
{[RP2006]\,979}  &	5 22 38.97	&	$-$67 55 09.2	&	15.881	&   15.402  &	0.7674	&	\ldots   & [1] \\
SMP\,LMC\,59     &	5 24 27.30	&	$-$70 22 23.9	&	\ldots 	&   19.058  &	0.2409	&	0.3176   & [2] \\
SMP\,LMC\,64     &	5 27 35.65	&	$-$69 08 56.3	&	16.455	&   15.702  &	8.3478	&	\ldots   & [2] \\
{[RP2006]\,695}  &  5 28 44.20  &   $-$69 54 22.5   &   16.923  &   15.334  &   23.6130 &   \ldots   & [1] \\
MGPN\,LMC\,55    &	5 31 09.07	&	$-$71 36 40.4	&	\ldots	&   20.055  &	1.0312	&	\ldots   & [3] \\
{[M94b]\,30}     &	5 31 35.18	&	$-$69 23 46.6	&	17.809	&   17.910  &	5.8833	&	\ldots   & [4] \\
\hline
\end{tabular}
\tablefoot{The table lists the names, coordinates (RA and Dec) for J2000, OGLE-III mean $V$ and $I$ magnitudes, primary and secondary period solution, and reference for the PN classification.}

\tablebib{[1]\,\citet{2006MNRAS.373..521R}, [2]\,\citet{2006A&A...456..451L}, [3]\,\citet{1998MNRAS.296..921M}, [4]\,\citet{2011A&A...531A.157M}.}

\end{table*}
%***************************************************************************************
%***************************************************************************************

%-------------------------------------------------------------
% ----------------------- OBSERVATIONS -----------------------
%-------------------------------------------------------------
%   Description of our observations.

\section{Observations}
\label{sec:observations}

   The OGLE survey \citep{1992AcA....42..253U}, which has been underway for more than 30 years, was originally designed to search for gravitational microlensing events. The project uses a dedicated 1.3 metre Warsaw Telescope located at Las Campanas Observatory, Chile. Currently, OGLE monitors the brightness of more than two billion point sources in the Milky Way and Magellanic Clouds. The OGLE photometric databases provide an excellent opportunity for researchers to examine the short- and long-term variability of many kinds of objects, including the central stars of planetary nebulae.
   
   OGLE uses $I$- and $V$-band filters closely reproducing the standard Johnson-Cousins system bands. The LMC photometry spans the range of $13<I<21.7$\,mag. The OGLE-III $I$-band filter differs significantly from the OGLE-IV $I$-band filter. \citet{2009A&A...496..813M} describes in detail the contamination of nebular emission. Nebular contamination can occur for the PNe with strong lines between 7000\,\AA\ and 9000\,\AA.

%-------------------------------------------------------------
% --------------------- DATA ANALYSIS ------------------------
%-------------------------------------------------------------
\section{Data analysis}
\label{sec:data_analysis}

We made a preliminary analysis of OGLE-III and/or OGLE-IV $I$-band photometry of 280 objects (see Tables~\ref{tab:pne_id_short} and A.1), using the Period04 software \citep{2005CoAst.146...53L}. We performed discrete Fourier transform (DFT) up to a frequency of 10\,${\rm c}\,{\rm d^{-1}}$ (cycles per day), corresponding to the period of 2.4 hours. This is close to the shortest period of a binary central star currently known of 2.3 hours \citep{2021MNRAS.506.5223J}. However, shorter periods, if present, would show up as aliases at lower frequencies. The OGLE observations have a typical cadence of about 1--3\,days and have seasonal gaps.

All the light curves and periodograms were visually inspected. If any significant peaks were found, we folded the time series with the derived period. Ellipsoidal variability produces two minima per orbital period, while irradiation variability produces one minimum per period. Consequently, the determination of the orbital period is not without ambiguity. In total 32 variables were found: 27 belong to the LMC, while 5 are foreground objects. The group of the LMC sources contains 18 objects (analysed in detail in Paper II); these objects show variability that cannot be attributed to binarity of CSPNe. The remaining nine objects showing a periodic variability (between 0.24 and 23.6\,days) are most likely genuine binary CSPNe. Of these nine objects, five were discovered only by \citet{2014MNRAS.438.2642R} and were not confirmed independently. The number of misidentified PNe in the sample by \citet{2014MNRAS.438.2642R} remains relatively large due to residuals present after sky-subtraction in the multi-object system used \citep{2011A&A...531A.157M} (see also Section 2 in this paper).

The remaining four objects have at least one independent detection using long-slit spectroscopic observations. None of the discovered binaries is located in the 30\,Dor star-forming region or other star-forming regions where \citet{2011A&A...531A.157M} found several PN mimics. The list of these objects is presented in Table~\ref{tab:list_binary_pne} along with the references for PN classification. The analysis for nine binary CSPNe was conducted independently with our own software routinely used for time-frequency analysis of photometry for classical pulsators \citep{2017MNRAS.468.4299S}(hereafter S17). The analysis using S17 follows a standard consecutive pre-whitening technique. Signiﬁcant periodicities are identiﬁed with the help of DFT and are modelled with a sine series fit. Residual data are inspected with DFT to detect additional lower-amplitude signals. Non-coherent signals are analysed and/or removed using time-dependent Fourier analysis \citep{1987ApJ...319..247K, 2015MNRAS.447.2348M}.

Both observations and common envelope modelling show that orbital periods of binary CSPNe cover the range from a few hours to a few tens of days \citep{2009A&A...496..813M}. We do not expect to find binary systems with longer orbital periods. They are expected to be very rare. 

We found 27 variable LMC objects in our sample of PNe. We split this sample into three groups. The first group contains nine objects with periods shorter than 30\,days, the second group contains objects with periods longer than 30\,days, and the third group is for objects with irregular or semi-regular variations. We consider the first group to be binary central star candidates. The second group shows much longer periods and may be explained by different type of objects (e.g. post-AGB or symbiotic stars). We do not consider these to be binary central stars since the amplitude of variability would be very small \citep{2008AJ....136..323D}. The third group may enclose the genuine PNe showing dust obscuration events \citep{2011A&A...528A..39M} or dust orbiting around the central star \citep{2008A&A...490L...7H}. The last two groups will be analysed using spectroscopic observations in Paper II.

One object showing periodic variability with a period shorter than 30\,days was discarded from the sample of binary central stars. [RP2006]\,1300 has a period of 2.34\,days, but does not show a spectrum typical for a PN (Paper II). The phased light curve reveals an eclipsing binary with an apsidal motion \citep{2013AcA....63..323P}. In addition, we checked the objects listed as variable in \citet{2007apn4.confE..69S}. We do not confirm that [RP2006]\,1304 is an eclipsing binary.

%-------------------------------------------------------------
% --------------- NOTES ON INDIVIDUAL OBJECTS ----------------
%-------------------------------------------------------------
\section{Notes on individual objects}
\label{sec:notes}

\begin{figure*}[h!]
   \resizebox{\textwidth}{!}
            {\includegraphics[]{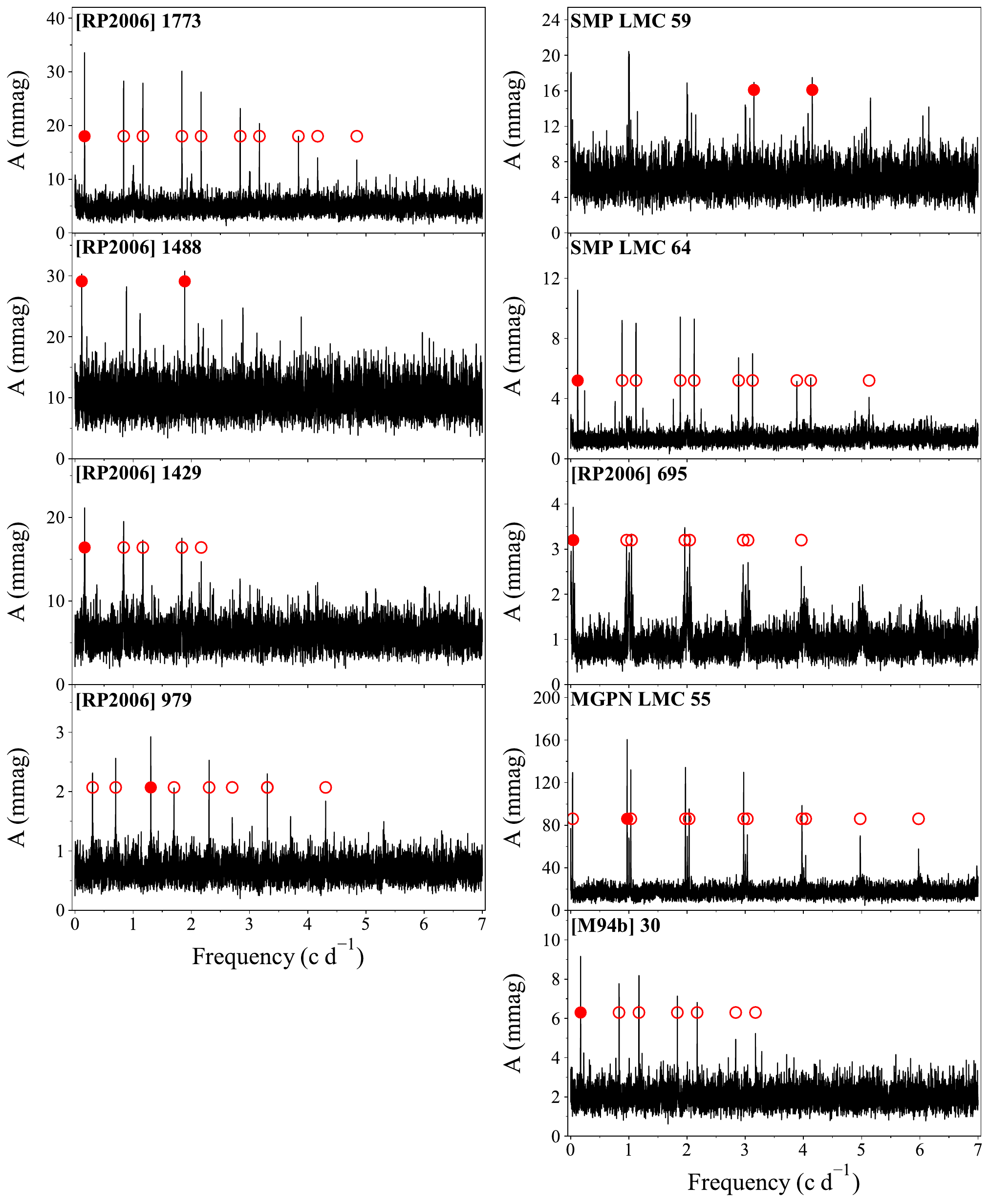}}
      \caption{Frequency spectra for binary CSPNe identified in the top left corner of each panel. The red filled circles correspond to the primary period, and in two cases the secondary period solution. The red open circles are the aliases.
              }
         \label{fig:periodograms}
   \end{figure*}

\begin{figure*}[h!]
   \centering % -> this comm was added by me
   \resizebox{0.8\textwidth}{!}
            {\includegraphics[]{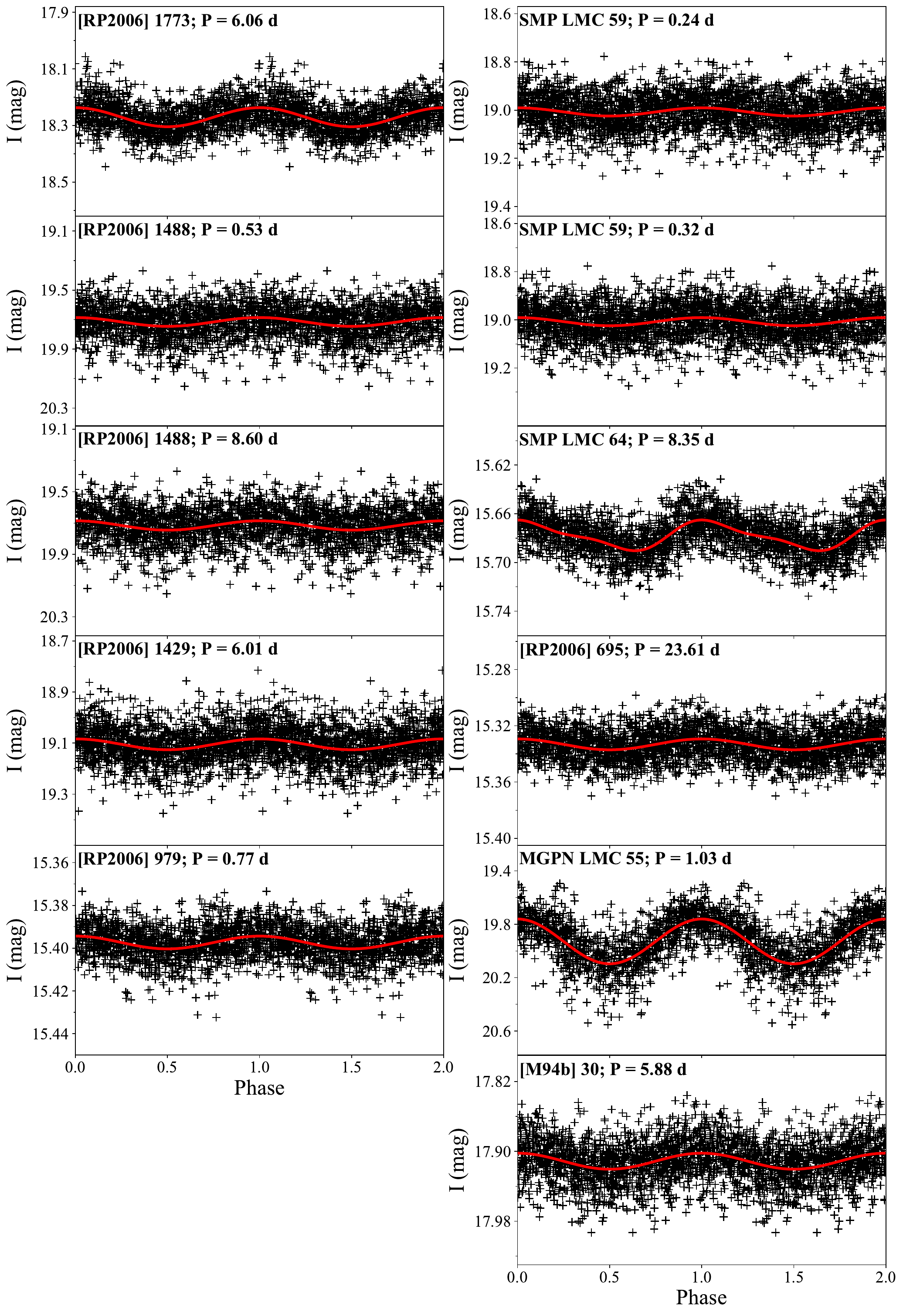}}
      \caption{Phased OGLE $I$-band light curves of the LMC 
      binary CSPNe identified in the top left corner of each panel. In the case of two PNe the light curve is also phased with the secondary period solution. The red line in each panel presents the fit to the data with the obtained period and amplitude.
              }
         \label{fig:phased_lc}
   \end{figure*}

{\bf[RP2006]\,1773} was classified as a True PN by \citet{2006MNRAS.373..521R}. Period04 analysis shows a certain variability with a period of 6.06\,days and a corresponding amplitude of 30\,mmag. Here and in the following, amplitude refers to a Fourier amplitude, the amplitude of a sine term fitted to the data to model the variability. The light curve of this object shows a non-physical variability for HJD above 6700\,days. With the S17 software we first removed a slow significant trend present in the data by applying a zeropoint correction to each of the observing seasons individually. Then, a significant peak ($S/N = 12.7$) is detected at $\nu = 0.164977\,{\rm c}\,{\rm d^{-1}}$ (Fig.~\ref{fig:periodograms}, filled red circle). The corresponding period and amplitude are 6.0615(2)\,days and 34\,mmag. The phased light curve is illustrated in Fig.~\ref{fig:phased_lc}. After pre-whitening, no other significant signals were detected in the frequency spectrum (the signal is coherent). 

{\bf [RP2006]\,1488} was classified as a True PN by \citet{2006MNRAS.373..521R}. The analysis with the S17 software reveals two peaks. These two signals are mutual daily aliases. In the frequency spectrum a `comb' of daily aliases is present. For the two highest peaks, with $S/N = 5.6$ and $S/N = 5.5$, we find, respectively, $P = 0.530097(4)\,{\rm d}$ ($A = 30\,{\rm mmag}$) and $P = 8.600(1)\,{\rm d}$ ($A = 30\,{\rm mmag}$). Since the difference in $S/N$ is too small, both solutions are likely. The detected signal is non-coherent, and we observe phase variation. The frequency spectrum is illustrated in Fig.~\ref{fig:periodograms}, and the data folded with periods corresponding to the two possibilities are plotted in Fig.~\ref{fig:phased_lc}. A preliminary analysis with Period04 leads to the same result.

{\bf [RP2006]\,1429} was classified as a True PN by \citet{2006MNRAS.373..521R}. Analysis with Period04 reveals the most likely period of 6.01\,days with a corresponding amplitude of 20\,mmag. With the S17 software, after detrending the data by adjusting the zero points of individual seasons, we find $S/N = 6.5$ for the highest peak in the frequency spectrum. The corresponding period and amplitude are 6.01296(4)\,days and 21\,mmag. The folded light curve is presented in Fig.~\ref{fig:phased_lc}. This variability is non-coherent; we observe clear phase variations. The frequency spectrum is illustrated in Fig.~\ref{fig:periodograms}. All other significant peaks are daily aliases. For the second highest peak we find $S/N = 6.0$ and a period of 1.19\,days. Based on the spectral window structure we judge that this solution is less probable. 

{\bf[RP2006]\,979} was classified as a True PN by \citet{2006MNRAS.373..521R}. With the S17 software, after removing the slow trend by adjusting zero points on a season-to-season basis, we detect ($S/N = 7.7$) a single periodicity with $P = 0.767396(5)\,\rm{d}$ and amplitude of only 3\,mmag. After pre-whitening, the unresolved remnant power remains in the frequency spectrum; the signal is non-coherent with significant phase variation. The frequency spectrum is shown in Fig.~\ref{fig:periodograms}; the folded light curve is plotted in Fig.~\ref{fig:phased_lc}. An independent Period04 analysis shows the two highest peaks in the frequency spectrum, which correspond to $P = 0.767402\,{\rm d}$ ($A = 3\,{\rm mmag}$) and $P = 1.429320\,{\rm d}$ ($A = 2.8\,{\rm mmag}$). Based on the spectral window structure we judge that the solution with $P\approx 1.43\,{\rm d}$ is less probable.

{\bf SMP\,LMC\,59} was classified as a PN by \citet{1978PASP...90..621S}, and its status as a PN was retained by \citet{2006MNRAS.373..521R}. The Hubble Space Telescope images reveal a complex quadrupolar morphology of the PN \citep{2001ApJ...548..727S}. In an analysis using the S17 software the trend was removed by adjusting zero points on a season-to-season basis. This time, however, residual significant power still remains at close-to-integer frequencies, which may correspond either to a remnant trend or to real variability with close-to-integer frequency. The ground-based data do not allow for verification. Otherwise, we detect significant power close to $4.15\,{\rm c}\,{\rm d^{-1}}$ ($S/N = 5.3$) and its daily alias at $3.15\,{\rm c}\,{\rm d^{-1}}$ ($S/N = 5.1$) (see the frequency spectrum in Fig.~\ref{fig:periodograms}). The difference in S/N is too small to decide which signal is real. The corresponding periods (determined after fitting sine terms to the data) are 0.2408737(8)\,days ($A = 17\,{\rm mmag}$) and 0.317580(1)\,days ($A = 17\,{\rm mmag}$). The signals are coherent. The light curves folded with the two periods are shown in Fig.~\ref{fig:phased_lc}. The Period04 analysis leads to the same results. 

{\bf SMP\,LMC\,64} is referred to as the youngest planetary nebula \citep{1991ApJ...374L..21D}, and is classified as a Possible PN by \citet{2006MNRAS.373..521R}. Its central star has reached the temperature of 31,500\,K. Its spectrum was obtained and analysed by \citet{2020A&A...642A..71H}. In an analysis using the S17 software, the strong trend present in the data was removed by adjusting zero points on a season-to-season basis. Then, the frequency spectrum (see Fig.~\ref{fig:periodograms}) is dominated by a strong ($S/N = 15.4$) peak. Its harmonic is also clearly visible ($S/N = 8.0$ after pre-whitening with the dominant signal). The corresponding period is 8.3481(3)\,days and the amplitude is 11\,mmag. The variability is non-coherent. Strong amplitude and phase variations are detected. Its light curve, with a clearly non-sinusoidal shape, is presented in Fig.~\ref{fig:phased_lc}. The nature of this variable is yet to be verified. The Period04 analysis leads to the same results.

{\bf[RP2006]\,695} was classified as a True PN by \citet{2006MNRAS.373..521R}. The variability of [RP2006]\,695 is dominated by a long-period non-coherent variability on timescales of a few hundred to a thousand days. Once the two strongest low-frequency signals were pre-whitened, the $S/N = 7.8$ signal was detected with a period of 23.613(5)\,days and an amplitude of 4\,mmag. The frequency spectrum is presented in Fig.~\ref{fig:periodograms}. The signature of long-period variability is still clearly visible (peaks at close-to-integer frequencies). The variability with a period of 23.6\,days is non-coherent with significant phase variation. The folded light curve is illustrated in Fig.~\ref{fig:phased_lc}. The Period04 analysis leads to the same results.

{\bf MGPN\,LMC\,55} was classified as a PN by \citet{1992A&AS...92..571M} using objective prism spectroscopy. However, this object was not analysed by \citet{2006MNRAS.373..521R}. In the analysis using the S17 software a strong signal of $S/N = 16.7$ dominates the frequency spectrum (Fig.~\ref{fig:periodograms}). The corresponding period and amplitude are 1.031215(3)\,days and 168\,mmag. No remnant power is detected after pre-whitening; the signal is coherent with constant amplitude and phase. A weak trend is detected, which can be removed with a low-order polynomial. The folded light curve is shown in Fig.~\ref{fig:phased_lc}. The Period04 analysis confirms the period of 1.03\,days with an amplitude of 161\,mmag.

{\bf [M94b]\,30} was classified as a PN by \citet{1994A&AS..103..235M}, and its status as a PN was retained by \citet{2006MNRAS.373..521R}. It is a possible bipolar PN \citep{2011A&A...531A.157M}. The analysis with the S17 software shows a single coherent periodicity of $P = 5.8833(3)\,{\rm d}$ and $A = 9\,{\rm mmag}$ ($S/N = 8.2$). The frequency spectrum is plotted in Fig.~\ref{fig:periodograms}, and the folded light curve is presented in Fig.~\ref{fig:phased_lc}. The analysis with Period04 leads to the same result.

%*************************************************************************************************************************
%*************************************************************************************************************************
%
%_____________________________________________________________
%                                             Two column Figure 
%-------------------------------------------------------------
% I added [ht] command to force LATEX to put the graph right here
% The valid options for the position specifier are:
% h : approximately here
% t : top of the page
% b : bottom of the page
% ! : to put emphasis on a specifier --> h! means "really here"
%
%
%*************************************************************************************************************************
%*************************************************************************************************************************

%*************************************************************************************************************************
%*************************************************************************************************************************
%
%_____________________________________________________________
%                                             Two column Figure 
%-------------------------------------------------------------
% I added [ht] command to force LATEX to put the graph right here
% The valid options for the position specifier are:
% h : approximately here
% t : top of the page
% b : bottom of the page
% ! : to put emphasis on a specifier --> h! means "really here"
%
%
%*************************************************************************************************************************
%*************************************************************************************************************************

%-------------------------------------------------------------
% ------------------------ DISCUSSION ------------------------
%-------------------------------------------------------------
\section{Discussion}
\label{sec:discussion}

We selected photometric variables with a period range that suggests binarity. We used similar approach to other photometric searches for binary central stars. The ultimate evidence of binarity would come from the radial velocity observations. The discovered binary central star candidates are listed in Table~\ref{tab:list_binary_pne}. The periods range from 0.24 to 23.6\,days.

All the [RP2006] objects in the sample except for [RP2006]\,1773 and SMP\,LMC\,64 show non-coherent effects or additional variations. The phased light curves of binary central stars can be complex, which can result in additional spectral components present in the periodograms. They can show a sinusoidal variability for the irradiation effect, two maxima per orbital period for ellipsoidal variability, eclipses, and other effects. In addition, the OGLE light curves may not be free from instrumental effects related to the presence of resolved or unresolved nebular contribution from free-free continuum and emission lines. The most obvious effect is that the observed amplitude of the stellar variability is suppressed. However, other types of variability cannot be excluded, for example variable stellar wind which can occur on timescales of hours and days (see e.g. \citet{2013MNRAS.430.2923H} or \citet{2014A&A...567A..15H}). 

The remaining PNe, [RP2006]\,1773, SMP\,LMC\,59, MGPN\,LMC\,55, and [M94b]\,30, show coherent clear signals, which makes them firm candidates for binary central stars. In addition, the SMP\,LMC\,59 nebula shows a complex morphology, which is consistent with the common envelope scenario. However, an intrinsic variability can be excluded only after confirmation of the binarity by radial velocity measurements.

The incidence rate of binary central stars in our sample is $3.3^{+2.6}_{-1.6}\%$ (9/273).\footnote{From the sample of 280 we discarded only seven foreground objects. The error bars were computed for binomial distribution and a confidence level of $\mathrm{p}=0.9$. We included, however, the 18 variable objects to be discussed in Paper II, which are the PNe or their mimics.} This is consistent with the incidence rate for the SMC computed by \citet{2014A&A...561A...8H} (1/45). However, the incidence rate for the LMC is significantly lower than the binary fraction found by \citet{2009A&A...496..813M} in the Galactic bulge. This comparison is based on a sample of 18 out of 139 true and likely PNe, which were selected after accounting for systematic effects and the exclusion of PNe located in fields that were largely insensitive to binary CSPN detection because of insufficient sampling. The fraction ratio is 4.5 (the range of 1.9 and 10.5 with a 95\% confidence level). The data used by \citet{2009A&A...496..813M} originates from the same survey and has similar quality.

One possible explanation for the lower binary fraction ratio in the LMC is that our search is more incomplete compared to the Galactic bulge survey. One factor that affects the completeness of our search is the greater distance to the LMC compared to the Galactic bulge: 50\,kpc to the LMC \citep{2019Natur.567..200P} compared to 8\,kpc to the Galactic bulge \citep{2019A&A...625L..10G}. This decreases the S/N for the LMC central stars. On the other hand, the greater distance is compensated by much lower extinction towards the LMC compared to the Galactic bulge.

We note that the two lowest amplitude detections in our sample, [RP2006]\,979 (3\,mmag) and [RP2006]\,695 (4 mmag), are reported in the brightest targets; the mean $I$-band magnitude is 15.402\,mag and 15.334\,mag, respectively. As the noise level in the data increases with increasing magnitudes, one may expect the variability of such a small amplitude to be missed in less bright targets. The detection limit increases with decreasing mean brightness, which is illustrated for example in figure 5 of S17 for the OGLE LMC data. While that study is focused on classical Cepheids, the detection limits are estimated after the pulsation signal is removed. We can conclude that signals of amplitude as low as 3-4\,mmag should be detected in targets of mean $I$-band magnitude up to about 16.5\,mag, but not in dimmer targets. Consequently, our inventory of binary PNe is not complete and the incidence rate of binary PNe should be regarded as a lower limit.

The completeness of the binary search can also be affected by more compact angular sizes of the LMC PNe. The contribution of the unresolved nebular continuum and line emission dilutes the signal from the central star.

It is difficult to correct the observed binary fraction for these effects quantitatively. However, it seems unlikely that observational biases could account for the binary fraction in the LMC being approximately five times smaller compared to the Milky Way. Our results contradict \citet{2019ApJ...875...61M}, who showed that the binary star fraction for close systems ($d < 10\,{\rm{AU}}$) is anticorrelated with metallicity for solar-type stars. They estimated that the close-binary fraction in the LMC is between 30 and 60\%.

Our results are in much better agreement with those of \citet{2012MNRAS.423.2764N}, who estimated the fraction of close binaries in the LMC to be 7--9\%, with the peak in the period distribution at about four\,days, based on the study of sequence E variables (close binary red giants, which may be the immediate precursors of PNe).

\begin{figure}[]
   \centering
   \includegraphics[width=\hsize]{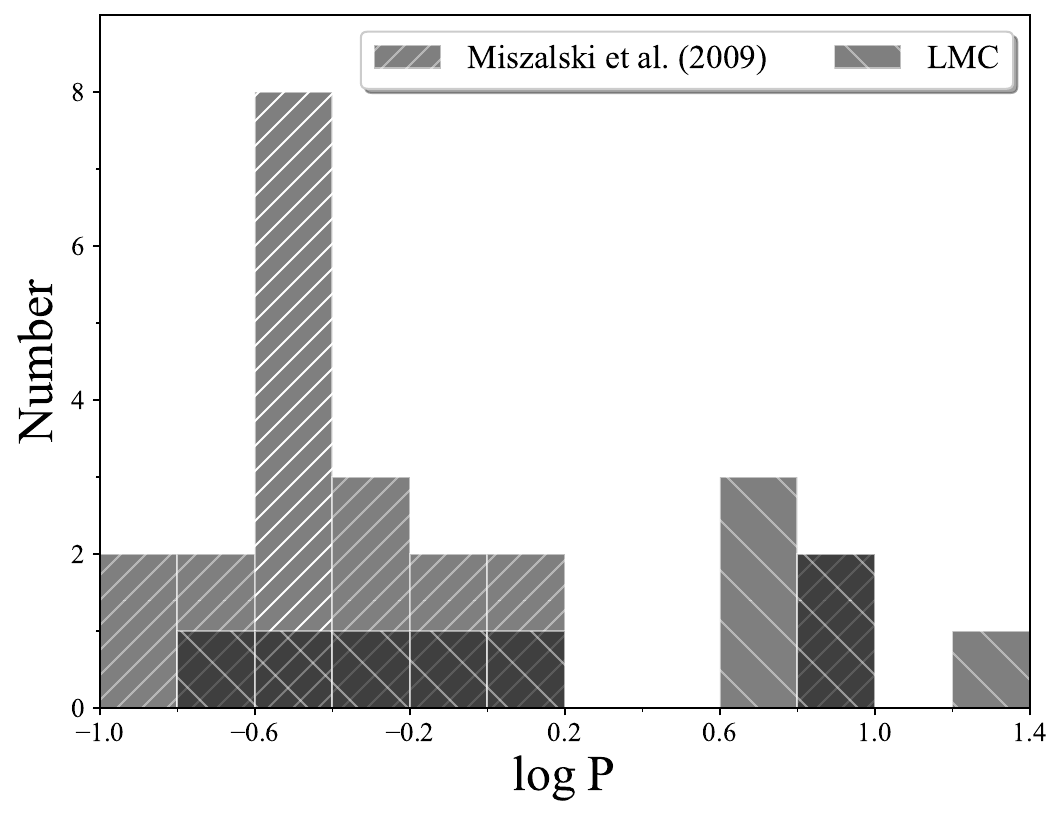}
      \caption{Comparison of the orbital period distribution of the new binaries in the LMC discovered in this paper, and in the Galactic bulge found by \citet{2009A&A...496..813M}.}
         \label{fig:period_distr_1}
   \end{figure}

\begin{figure}[]
   \centering
   \includegraphics[width=\hsize]{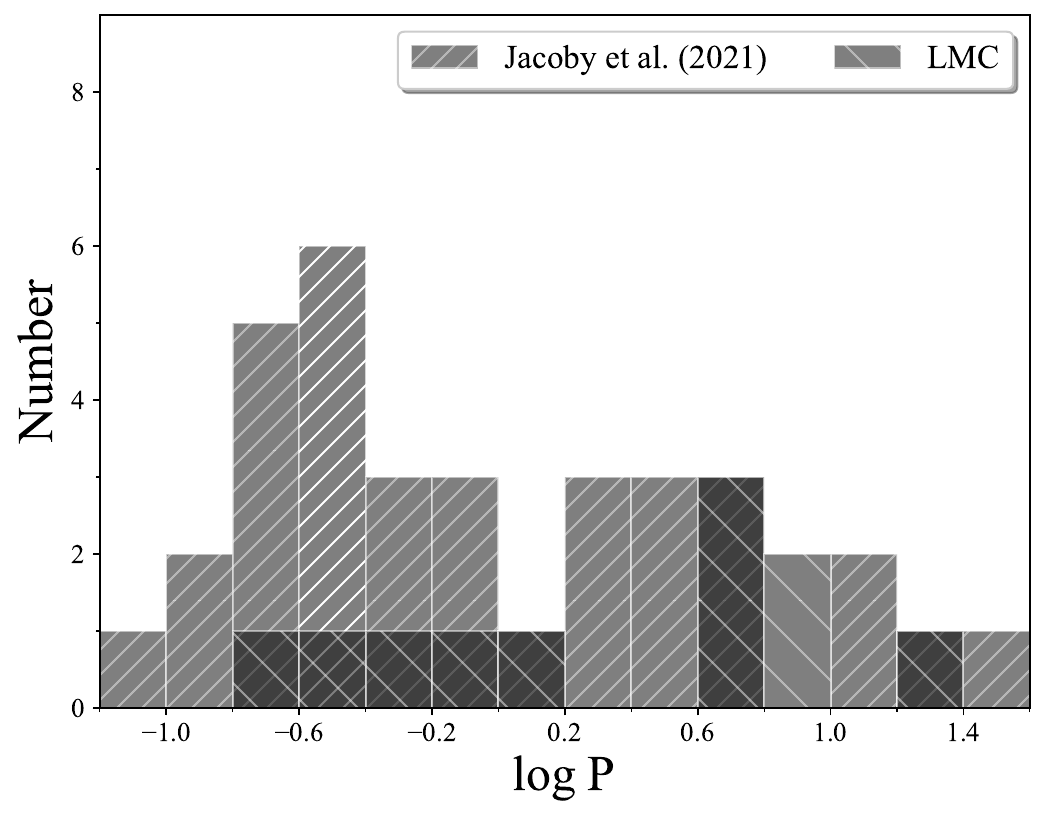}
      \caption{Comparison of the orbital period distribution of the new binaries in the LMC discovered in this paper, and in the Milky Way found by \citet{2021MNRAS.506.5223J}.}
         \label{fig:period_distr_2}
   \end{figure}

Figure~\ref{fig:period_distr_1} depicts the period distributions of our new sample, and the 21 known binaries from the Galactic bulge \citep{2009A&A...496..813M} with a chosen bin of $\Delta\log{P}=0.2$. Our sample contains two objects ([RP2006]\,1488 and SMP\,LMC\,59) with a two-period solution. Both periods are taken into account in the period distribution. The shape of the period distributions seems to be different. In the case of the Galactic bulge sample the peak around $\log{P} = -0.5$ is clearly visible, whereas in the LMC the distribution of periods does not show a well-defined maximum. However, the number of objects in the LMC is too low to compare the shape of the two period distributions. Figure~\ref{fig:period_distr_2} shows the period distribution of our new sample, and 34 known Galactic binaries from \citet{2021MNRAS.506.5223J} with a chosen bin of $\Delta\log{P}=0.2$. Two objects from the Galactic sample are not included in the histogram because their variability can be caused, for example, by stellar winds \citep{2021MNRAS.506.5223J}: PN\,H\,2-48 with a period of 4.1\,days and PN\,Th\,3-14 with a period of 1.15\,day. Again, the shape of the period distributions seems to be different, with a lack of a well-defined maximum in the case of the LMC sample. Interestingly, \citet{2009A&A...496..813M} obtained the maximum of binary period distribution at $\log{P} = -0.5$ corresponding to eight hours, while \citet{2021MNRAS.506.5223J} used space-based photometry and derived much wider period distribution with the periods ranging from 2.3\,h to 30\,days.

%***************************************************************************************
%***************************************************************************************
%_____________________________________________________________
%                 A figure as large as the width of the column
%-------------------------------------------------------------
% I used [H] command in the incl. graphics and \package{float} in the preambule
% to force Latex to put the graph where I wanted to put it.
%
%***************************************************************************************
%***************************************************************************************  

%***************************************************************************************
%***************************************************************************************
%_____________________________________________________________
%                 A figure as large as the width of the column
%-------------------------------------------------------------
% I used [H] command in the incl. graphics and \package{float} in the preambule
% to force Latex to put the graph where I wanted to put it.
%
%***************************************************************************************
%***************************************************************************************  

%-------------------------------------------------------------
% ------------------------- SUMMARY --------------------------
%-------------------------------------------------------------
\section{Summary}
\label{sec:summary}

We have analysed the light curve variability of 280 planetary nebulae from the Large Magellanic Cloud using photometry at $I$-band from the OGLE-III and OGLE-IV surveys. To search for periodic variables we used Period04 and our own software. We find nine periodic central stars of PNe with period ranging from 0.24 up to 23.6\,days, which we consider to be binary PNe central star candidates. The real nature of this variability may be confirmed by radial velocity measurements. In the best scenario, our investigations have increased the number of known extragalactic binary PNe to ten.

The period range and its distribution in the LMC seems to be different to that found by \citet{2009A&A...496..813M} towards the Galactic bulge, and by \citet{2021MNRAS.506.5223J} in the Milky Way in general, but the number of the LMC objects is too low to draw a conclusion. However, the fraction of binary PNe determined to be $3.3^{+2.6}_{-1.6}\%$ for the LMC PNe is much lower than that established for the Milky Way, but it is uncorrected for incompleteness. On the other hand, the lower fraction may result from observational selection since the amplitude of variations due to the great distance to the LMC may be insufficient. The effect of the greater distance is, however, at least partially mitigated by lower extinction to the LMC PNe. Thus, the suggested lower binary fraction seems to be real. Given that solar-type binary systems are more frequent in lower metallicity \citep{2019ApJ...875...61M}, they must be much less effective in producing close binaries through the common envelope than their higher metallicity counterparts. This suggests that systems that experience common envelope preferentially either end up at larger separations and remain undetected or end up as mergers. The fate of the system experiencing a common envelope phase is not yet well constrained theoretically. It depends on the binding energy of the envelope of the AGB star at the time of the contact and the mass ratio of the components \citep{2011MNRAS.411.2277D}. Further study of close binaries in the PNe in different environments can help us to constrain the physics of common envelope interactions.

\begin{acknowledgements}
      MG acknowledges support from the National Science Centre grant 2014/15/N/ST9/04629. MH acknowledges financial support from the NCN grant 2016/23/B/ST9/01653. RSz acknowledges support from the NCN grant 2016/21/B/ST9/01626. IS acknowledges support from the National Science Centre, Poland, grant OPUS no. 2022/45/B/ST9/00243. For the purpose of Open Access, the author has applied a CC-BY public copyright licence to any Author Accepted Manuscript (AAM) version arising from this submission. 
      RSm is supported by the Polish National Science Centre, SONATA BIS grant, 2018/30/E/ST9/00598.
      We have made extensive use of the SIMBAD and Vizier databases 
      operated at the Centre de Donn\'ees Astronomiques de Strasbourg. 
      This research has made use of "Aladin sky atlas" developed at CDS, Strasbourg Observatory, France.
      The OGLE project has received funding from the European 
      Research Council under the European Community's Seventh 
      Framework Programme (FP7/2007-2013)/ERC grant agreement 
      no. 246678.
      
      This work has made use of data from the European Space Agency (ESA) mission {\it Gaia} (\url{https://www.cosmos.esa.int/gaia}), processed by the {\it Gaia} Data Processing and Analysis Consortium (DPAC, \url{https://www.cosmos.esa.int/web/gaia/dpac/consortium}). Funding for the DPAC has been provided by national institutions, in particular the institutions participating in the {\it Gaia} Multilateral Agreement.

\end{acknowledgements}


%-------------------------------------------------------------
% ----------------- THE BIBLIOGRAPHY - BIBTEX ----------------
%-------------------------------------------------------------
\bibliographystyle{aa}
\bibliography{aa}
%
%_____________________________________

%-------------------------------------------------------------
%------------------------- APPENDICES ------------------------
%-------------------------------------------------------------

\begin{appendix}
{

%-------------------------------------------------------------
% ---------------- APPENDIX - IDENTIFIED PNe -----------------
%-------------------------------------------------------------
\section{Identified PNe in the LMC} %Second online appendix
\label{sec:app_pne_id}

   In Section~\ref{sec:pne_sample} we present Table~\ref{tab:pne_id_short} with the list of the LMC PNe listed in \citet{2014MNRAS.438.2642R} identified in the OGLE-III and OGLE-IV fields, but only for the first five objects from the entire sample, which consists of 280 objects. Here, we present the full list of identified objects. Table A.1 is only available at the CDS.
   
}
\end{appendix}

\end{document}